\documentclass[twocolumn,nolinenumbers]{aastex631}
\usepackage{graphicx}	% Including Figure~files
\usepackage{amsmath}	% Advanced maths commands
\usepackage{amssymb}	

\include{defns}
\newcommand{\ngc}{{NGC~7469}}

\usepackage{comment}

\begin{document}
\title{Testing X-ray Reprocessing and Mapping the Soft Excess of NGC~7469 with NICER}
%First Author:
\author[0000-0003-1183-1574]{Ethan R. Partington}
\affiliation{Department of Physics and Astronomy, Wayne State University, 666 W.\ Hancock St, Detroit, MI, 48201, USA}
\affiliation{Institute of Astrophysics, FORTH, GR-71110, Heraklion, Greece}

\author[0000-0002-8294-9281]{Edward M.\ Cackett}
\affiliation{Department of Physics and Astronomy, Wayne State University, 666 W.\ Hancock St, Detroit, MI, 48201, USA}

\author[0000-0001-8598-1482]{Rick Edelson} 
\affiliation{Eureka Scientific Inc., 2452 Delmer St. Suite 100, Oakland, CA 94602, USA}

\author[0000-0003-1728-0304]{Keith Horne}
\affiliation{SUPA School of Physics and Astronomy, North Haugh, St.~Andrews, KY16~9SS, Scotland, UK}

\author[0000-0001-8475-8027]{Jake A. Miller}
\affiliation{Department of Physics and Astronomy, Wayne State University, 666 W.\ Hancock St, Detroit, MI, 48201, USA}
\affiliation{Mitchell Institute for Fundamental Physics \& Astronomy, Texas A\&M University, 576
University Dr., College Station, TX 77843, USA}

\author[0000-0002-3026-0562]{Aaron J.\ Barth}
\affiliation{Department of Physics and Astronomy, 4129 Frederick Reines Hall, University of California, Irvine, CA, 92697-4575, USA}

\author[0000-0001-9092-8619]{Jonathan Gelbord}
\affiliation{Spectral Sciences Inc., 4 Fourth Ave., Burlington, MA 01803, USA}

\author[0000-0002-6733-5556]{Juan V.\ Hern\'{a}ndez Santisteban}
\affiliation{SUPA School of Physics and Astronomy, North Haugh, St.~Andrews, KY16~9SS, Scotland, UK}

\begin{abstract} 
We present an X-ray/UV reverberation analysis of NGC 7469 across 210 days, using daily NICER observations with contemporaneous monitoring by Swift UVOT+XRT. We model the X-ray spectrum with a power law continuum and a soft excess during each NICER epoch. These emission sources demonstrate correlated flux variability with a lag consistent with zero days. We find that the power law emission is consistent with a compact X-ray corona, and that the soft excess can be explained by reflected coronal emission from the inner accretion disk.
We test the relationship between changes in the flux of the X-ray corona and the UVW2 continuum, finding strong correlation and a negative X-ray lag of less than one day. This is consistent with a scenario in which the X-ray corona drives the UVW2 light curve through thermal reprocessing.   
\end{abstract}

\section{Introduction}
AGN can heavily impact their host galaxy through radiative feedback and outflows, requiring comprehensive knowledge of their accretion mechanisms to understand their role in galactic evolution. This includes an assessment of the heating mechanism of the accretion disk, both through internal viscous interactions and through the reprocessing of ionizing radiation produced in the extreme environment close to the black hole. We aim to test the origin of fast X-ray and UV variability in the nearby, X-ray bright Seyfert 1 AGN NGC~7469 with $M_\mathrm{BH}=9 \times 10^6 M_\odot$ (from the database in \citealt{Bentz15}, using $\langle f \rangle=4.3$ from \citealt{Grier_2013}) at $z=0.0163$ \citep{Springob_2005}. The AGN shows weak X-ray obscuration which is stable on timescales of decades (\citealt{Peretz_2018}, \citealt{Grafton-Waters_2020}). This will allow us to measure the intrinsic X-ray variability without complication from variable outflows, such as those in Mrk~817 \citep{partington23}.

Prince et al. (in prep.) estimates the bolometric luminosity of \ngc\ to be on the order of 10\% of the Eddington luminosity. For black holes accreting at this rate, the accretion disk is typically modeled as a geometrically thin, optically thick disk which emits thermal radiation as a blackbody \citep{Shakura1973}. The temperature should increase at smaller radii (measured relative to the central black hole). Given the mass of \ngc, thermal emission is expected to peak in the UV at a few hundred gravitational radii ($R_\mathrm{G}=GM_\mathrm{BH}/c^2$). 

At smaller radii, the plasma becomes too hot to thermalize, forming a ``corona" of hot electrons. The electrons are cooled through inverse Compton scattering with the disk photons, releasing X-rays with a power law energy distribution. An increase in the available seed photons, through the brightening of the disk, will cool the corona. More soft X-rays will be produced, changing the shape of the power law spectrum (\citealt{SUNYAEV79}, \citealt{Haardt_1991}). X-rays traveling towards the disk will then be absorbed, increasing the disk's local temperature. This scenario was shown to be likely for \ngc\ by \cite{Nandra_2000} and \cite{Petrucci_2004} through spectral modeling of RXTE observations, integrated daily in a campaign lasting for 30 days. Analysis of this dataset by \cite{Papadakis_2001} showed that the variability occurs in the soft X-rays before the hard X-rays, consistent with thermal Comptonization in a compact corona.

In AGN, the X-ray flux is highly variable on timescales of a few days. In the thermal reprocessing scenario, the disk is expected to produce UV/optical light curves which track variations in the X-rays. In a given UV/optical band, features in the light curve should lag behind the X-rays by the light travel time between the X-ray corona and the disk radius where the majority of the in-band flux is emitted (e.g., \citealt{Cackett2007}). The technique of measuring a time lag between variability in spatially distinct emission regions of the AGN to obtain their separation distance is known as ``reverberation mapping" (see the review by \citealt{cackett21} and references therein). 

The size of the X-ray corona can be measured by comparing variations in the incident power law continuum and the Fe K$\alpha$ reflection line near 6.4 keV, which has broad tails caused by the extreme rotation of the inner disk. \cite{kara16} measured the broad Fe K$\alpha$ lag in \ngc, estimating the corona's size to be on the order of $10 R_\mathrm{G}$. The soft excess, an additional emission source which emits primarily below 2 keV, may also be a reflected signature of the power law photons, which are Compton down-scattered off of the inner disk atmosphere and gravitationally redshifted (\citealt{Crummy_2006}, \citealt{Rozanska_2008}, \citealt{Garcia_2014}).

In many other AGN, the observed hard X-ray/UV correlation is much weaker than expected for the standard X-ray reprocessing scenario (e.g. \citealt{mehdipour11}, \citealt{Edelson2015}, \citeyear{edelson2017}, \citealt{Gardner2017}, \citealt{Partington_2024}). This suggests that the accretion disk geometry is sometimes more complicated than the standard \cite{Shakura1973} picture. A popular explanation for this behavior is that the inner disk fails to maintain thermal equilibrium and expands vertically. An optically thick layer of electrons known as the ``warm corona" may form, which may disrupt direct X-ray reprocessing and contribute to the emission of the soft excess (\citealt{mehdipour11}, \citealt{Done12}). Alternatively, the corona may be dynamic \citep{Panagiotou_2022_xraycorrel}. By modeling the X-ray spectra of a large sample of AGN using eROSITA, \cite{waddell_2023} found that models of a warm corona and relativistic reflection were each preferred for different AGN, indicating that the origin of the soft excess is not uniform. 

In some AGN including \ngc\ (\citealt{Kumari_2023}), the X-ray/UV correlation is reduced by rapid variability in the X-rays on timescales of a day or less which are not fully reproduced in the UVW2 light curve. Slow variability on timescales of several days or more is also commonly observed in both bands (see \citealt{Pahari_2020} for an example in \ngc). By averaging the light curves over longer timescales than the sampling cadence, these trends can be isolated. These slow variations can sometimes have a stronger correlation than the raw light curves, such as in Fairall~9 \citep{Partington_2024}. \cite{Hagen_2024} proposes that these slow variations are introduced through some process in the disk, either from intrinsic variability which spans a large range of radii (e.g. \citealt{Arevalo_2006}, \citealt{Arevalo_2008} \citealt{Neustadt_2022}), or from instabilities at the truncated inner accretion flow. This would modulate the seed photons available to the corona for Comptonization, introducing a slow trend in the X-rays which lags behind the UV. Additional intrinsic variability in the corona would produce the fast signal. This would then be reprocessed in the UV, lagging behind the X-rays. 

\begin{figure*}[th!]
    \centering
    \includegraphics[width=0.9\textwidth]{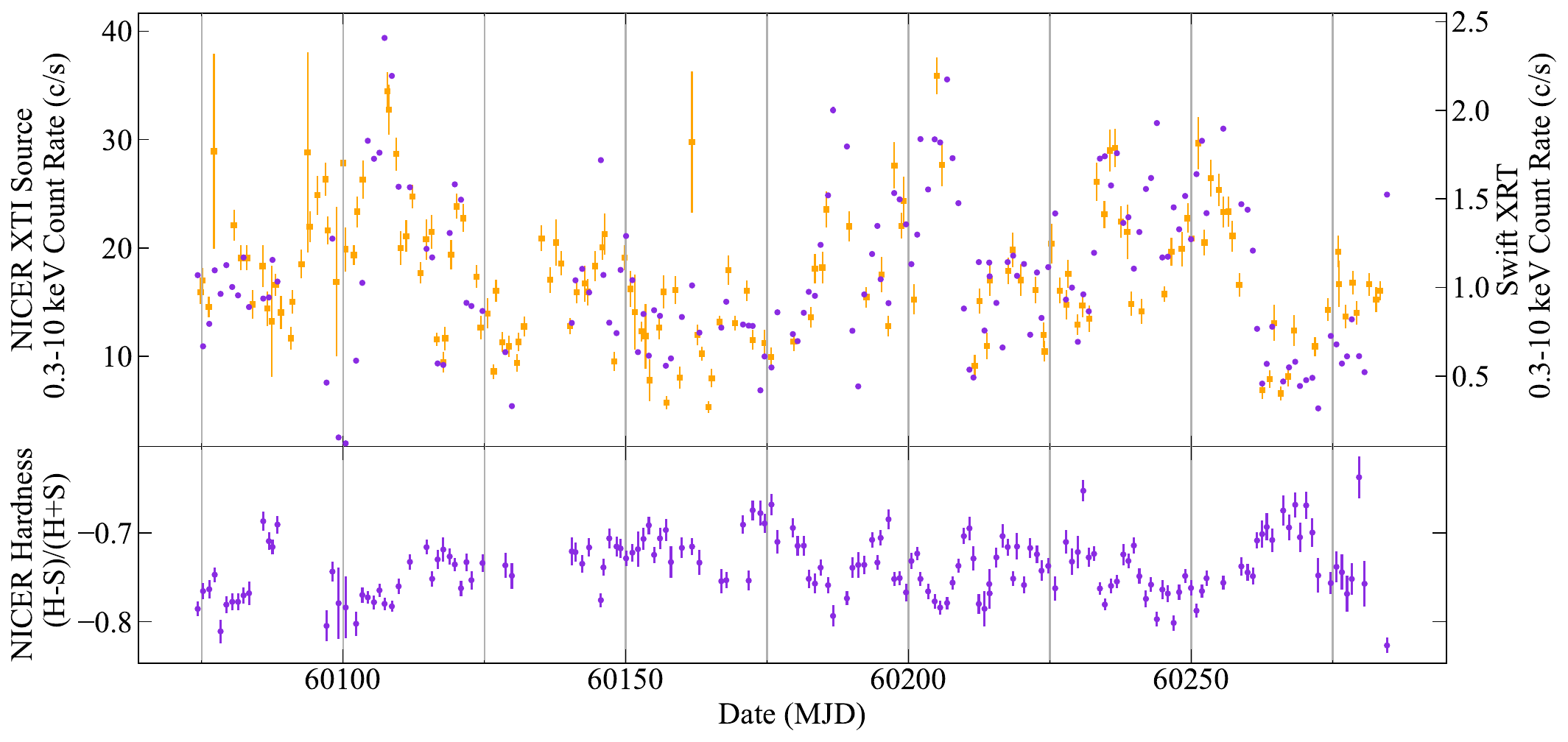} 
    \caption{\textbf{\textit{Top}}: X-ray count rate light curves, measured from 0.3--10 keV with NICER XTI (purple circles) and Swift XRT (orange squares). The vertical axes are scaled by a factor of 16.5 to account for the instrumental differences in count rate sensitivity. Error bars representing $1\sigma$ uncertainty are shown for every data point, but for NICER they are always smaller than the width of the point itself. \textbf{\textit{Bottom}}: The hardness ratio for NICER observations, calculated such that the hard band $H$ covers 2--10 keV and the soft band $S$ covers 0.3--2 keV.}  
\label{fig:lcCompSwiftNICERXrays}
\end{figure*}

Our long-term monitoring of \ngc\ will test the X-ray/UV reprocessing scenario through spectral modeling of isolated emission components. We also aim to reveal if the X-ray soft excess varies in the same manner as the X-ray corona, indicative of inner disk reverberation, or whether alternative scenarios are required. Daily spectra are needed for this, and Swift and NICER are the only observatories with the required scheduling capability. Only NICER is sensitive to changes in the spectral parameters on these timescales, due to its nearly $\sim20\times$ higher throughput than Swift (see Figure~\ref{fig:lcCompSwiftNICERXrays}). We describe our data reduction method in Section~\ref{sec:Data} and our analysis in Section~\ref{sec:Analysis}. We discuss our results in Section~\ref{sec:Discussion} and share the main conclusions of our work in Section~\ref{sec:Conclusion}. We compare the instrumental sensitivity of Swift and NICER to the parameters in our spectral model in Appendix~\ref{sec:swiftappendix}.

\section{Observations and Data Reduction}
\label{sec:Data}
We process daily NICER observations using \textsc{HEASOFT} version 6.35 \citep{Blackburn_1995} and \textsc{CALDB} version xti20240206, using the standard filtering settings on the \textsc{NICERL2} task from the \textsc{FTOOLS} software package. This includes the filtering of ``noise ringers," a non-astrophysical current in the NICER electronics. This signal persists for up to $110\mu\mathrm{s}$ after exposure to high optical photon fluxes, which are falsely counted as soft X-ray photons. The accumulation of optical light in the silicon X-ray detectors is known as ``optical loading." It typically produces a Gaussian spectral shape centered below 0.3 keV. The presence of noise ringers can produce an additional spectral feature at energies up to 0.6 keV. Without filtering, this can affect soft X-ray spectral analysis within the source sensitivity band of 0.3--10 keV. An ``optical light leak" began on 22 May 2023 (MJD 60086), caused by a new hole in the instrument's optical light filter. This has greatly increased the effect of optical loading during the ISS day. 

We restrict our analysis to observations taken between 10 May 2023 and 06 Dec 2023 (MJD 60074--60284). Observations from MJD 60285--60307 are part of the planned campaign but were taken during the ISS day, resulting in filtered exposure times of zero seconds due to the optical light leak. Although there are an additional seven observations between MJD 60316--60321 with nonzero exposure times, the large gap in the light curve from including these epochs significantly degrades our cross-correlation analysis performed in Section~\ref{sec:crosscorrelation}. We also restrict our sample to epochs with exposure time $<$ 300 s to ensure sensitivity to our source parameters, resulting in 176 epochs. 

We additionally exclude one observation which exhibits extreme optical loading even after filtering, indicated by a count rate of 1,055 c~s$^{-1}$ in the 0.3--1 keV band. Without this epoch, the median 0.3--1 keV count rate is 10 c~s$^{-1}$ with a range of 1--24 c~s$^{-1}$. High count rates from 10--12 keV indicate particle background flares which extend into the source sensitivity band of 0.3–10 keV, preventing an accurate measurement of the shape and flux of the AGN power law spectrum. We thus exclude 11 observations with 10--12 keV band count rates $>$ 1 c~s$^{-1}$, corresponding to 6\% of the epochs. In the remaining sample, the median 10--12 keV count rate is 0.09 c~s$^{-1}$ and the mean is 0.15 c~s$^{-1}$. We restrict our analysis to these 164 epochs, presented in Figure~\ref{fig:lcCompSwiftNICERXrays} and Table~\ref{table:nicerctrthardness}.
\begin{deluxetable}{ccc}[t]
\tablewidth{0pt}
\tablecaption{\label{table:nicerctrthardness}NICER Count Rate and Hardness}
\tablehead{
\colhead{Obs. Date} &
\colhead{Source Count Rate} &
\colhead{Hardness Ratio}\\ 
\colhead{(MJD)} &
\colhead{0.3--10 keV} &
\colhead{H=2--10 keV}\\
\colhead{} &
\colhead{(counts~s$^{-1}$)} &
\colhead{S=0.3--2 keV}
}
\startdata
60074.34 & $17.5\pm0.1$ & $-0.786\pm0.008$\\ 
60075.28 & $10.9\pm0.1$ & $-0.766\pm0.009$\\ 
60076.38 & $13.0\pm0.1$ & $-0.763\pm0.011$\\ 
60077.34 & $17.9\pm0.1$ & $-0.747\pm0.008$\\ 
60078.38 & $15.8\pm0.2$ & $-0.811\pm0.013$\\ 
... & ... & ... 
\enddata
\tablecomments{Source count rates are calculated by subtracting the background count rate from the total count rate for each observation. Listed uncertainties represent the 68\% confidence interval ($1\sigma$). The hardness ratio is calculated as $(H-S)/(H+S)$. The full table is available in machine-readable form.}
\end{deluxetable}

\begin{figure*}[t]
    \centering
    \includegraphics[width=0.9\textwidth]{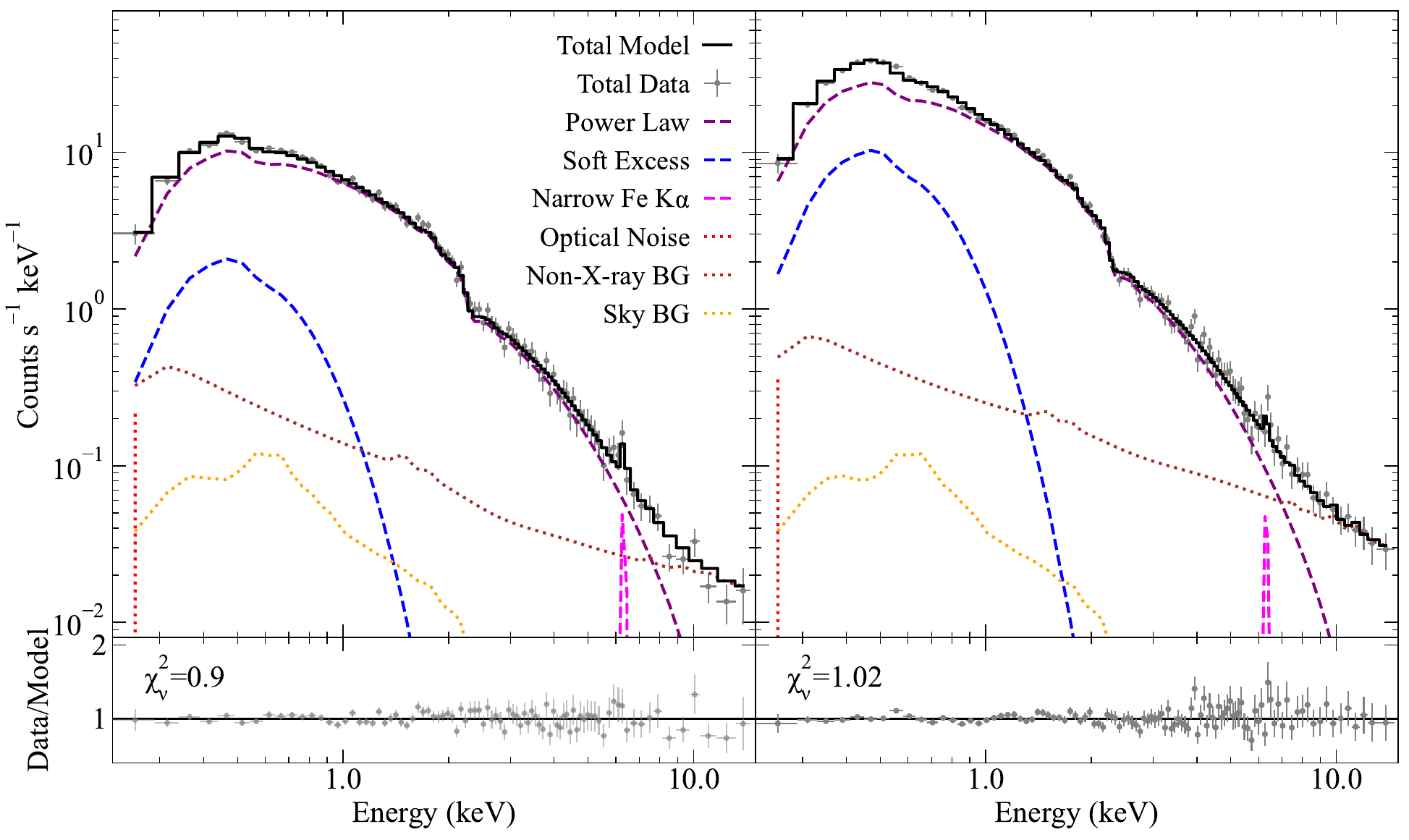} 
    \caption{Count rate NICER spectra of \ngc\ from two epochs characterized by a bright (right) and faint (left) soft excess. The total observed spectrum including the source and background counts is shown as grey points, and the total model is shown as a black line. The source model components of a power law (purple), soft excess (blue) and narrow Fe K$\alpha$ reflection line (pink) are plotted as dashed lines. Dotted lines represent the optical noise (red), X-ray sky (yellow), and non-X-ray (brown) background components of the SCORPEON model. The ratio between the data and model for each energy bin is plotted as gray points in the panel below each spectrum. \textbf{\textit{Left}}: An observation taken on MJD 60179 (ObsID 6714020701) while the soft excess flux is low ($\log(\Phi_{\mathrm{SE}})=-11.93\substack{+0.08\\ -0.16}$). All fluxes are reported in units of erg~cm$^{-2}$~s$^{-1}$. The flux of the X-ray power law is $\log(\Phi_{\mathrm{PL}})=-10.527\substack{+0.007\\ -0.006}$ and $\Gamma=1.72\pm0.03$. \textbf{\textit{Right}}: An observation from MJD 60204 (ObsID 6714023201), taken near a peak in the soft excess light curve ($\log(\Phi_{\mathrm{SE}})=-11.24\pm0.06$). The X-ray power law is softer ($\Gamma=1.94\substack{+0.03\\ -0.04}$) and brighter ($\log(\Phi_{\mathrm{PL}})=-10.231\pm0.005$) than in the observation on MJD 60179.}
\label{fig:spec_comp}
\end{figure*}

The SCORPEON background model (Figure~\ref{fig:spec_comp}) is currently the only NICER background characterization tool able to characterize the variable shape of the optical noise peak. As such, it is recommended by the NICER team for the analysis of any soft X-ray spectra taken after MJD 60086. SCORPEON was also demonstrated to be the optimal background tool for analysis of AGN prior to the light leak in \cite{Partington_2024}. Our analysis of \ngc\ is thus conducted using the SCORPEON background, produced with the \textsc{NICERL3-spect FTOOL}. Source count rates after background subtraction range from 2.0--39.4 counts~s$^{-1}$ (median 16.7 c~s$^{-1}$) in the 0.3--10 keV band.

We produce a UVW2 band light curve using data from Swift UVOT, following the methodology of \cite{Hernandez2020}. The light curve covers MJD 60074--60286 with an average observing cadence of $1.35\times$ per day. 

Swift experienced a pointing stability anomaly that affects all data taken after MJD 60163. The resulting jitter frequently causes the image PSF to be smeared beyond the standard UVOT measurement extraction region. To minimize any excess aperture losses, we have doubled the extraction region radius from $5\arcsec$ to $10\arcsec$ and excluded one observation in which the smeared PSF extends beyond $\pm 10\arcsec$. The smearing also degrades the estimation of the degree of pileup, reducing the accuracy of the UVOT coincidence loss correction factors and biasing these towards low values. Consequently, the UVOT fluxes from MJD 60164 onward reported here are noisier and systematically low and should be considered preliminary.

We note, however, that in the case of NGC 7469 in UVW2 the jitter imparts only a few percent effect upon the flux measurements, much smaller than the amplitude of UV variability, so features and trends in the light curve still provide meaningful constraints to cross-correlation analyses.
Swift XRT observes simultaneously with UVOT in the X-rays from 0.3--10 keV, and we generate a light curve and a spectrum for each epoch using the XRT product builder (\citealt{evans07}, \citealt{evans09}). Trends in the X-ray light curve are generally consistent between Swift and NICER (see Fig.~\ref{fig:lcCompSwiftNICERXrays}).

We take photometric measurements of \ngc\ in the \textit{g} band from MJD 60074--60326 using the Dan Zowada Memorial Observatory and the global Las Cumbres Observatory (LCO) telescope network, following the observing procedures of \cite{miller_2023} for Zowada and \cite{Hernandez2020} for LCO. We combine multiple observations taken by a single telescope during the same day into one epoch. To isolate erroneous photometric measurements while preserving variability intrinsic to the AGN, we exclude 23 epochs with a measured flux which differs from the median flux in a 7-day window by more than $20\sigma$, where $\sigma$ is the uncertainty in measured flux for that epoch. The average observing cadence of the remaining 383 epochs, including all telescopes, is $1.51\times$ per day. We combine and intercalibrate the light curves from each telescope using the CALI pipeline \citep{Li2014} to produce the final light curve used for interband cross-correlation analysis in Section~\ref{sec:crosscorrelation}. 

\section{Analysis}
\label{sec:Analysis}
\subsection{X-ray Spectral Model}
\label{sec:specparam}

We fit the NICER X-ray spectrum from each epoch in XSPEC \textsc{v12.15.0} \citep{arnaud96} from 0.22--15 keV. We use the SCORPEON background model (see \citealt{partington23} and \citealt{Partington_2024} for a detailed discussion and comparison of NICER background tools for AGN analysis). We fix all parameters of the X-ray sky background to their default SCORPEON values, holding them constant throughout the campaign. Spectra are grouped following the optimal binning protocol of \cite{kaastra16} with a minimum of 25 counts per bin, using the \textsc{FTGROUPPHA} task from the \textsc{FTOOLS} software package. 

The source model is shown in Figure~\ref{fig:spec_comp} and Table~\ref{table:modelparams}. We apply fixed Galactic absorption of $N_\mathrm{H}=0.0448 \times 10^{22}$ cm$^{-2}$, calculated using the \textsc{FTOOLS} task \textsc{nH}. We include a Gaussian profile representing the narrow Fe K$\alpha$ reflection line at rest energy 6.4 keV and redshift $z=0.0163$ \citep{Springob_2005}. The weak obscuration from a warm absorber present in XMM RGS spectra \citep{Grafton-Waters_2020} is not statistically required in our model, given the limited spectral resolution of NICER. Similarly, we do not significantly detect the broad Fe K$\alpha$ line from inner disk reflection observed with XMM PN and NuSTAR \citep{Ogawa_2019}. The shape and flux of the soft excess alone is insufficient to constrain reflection parameters such as the black hole spin, disk inclination, and disk ionization. We thus measure the shape and flux of the soft excess using a source-agnostic blackbody model with variable temperature $kT$ (e.g. \citealt{Crummy_2006}). The flux of each source model component from 0.3--10 keV is calculated using \textsc{cflux}. 

We compare power law models with and without a cutoff energy ($E_\mathrm{cut}$). During this test, we fit each spectrum using the following free parameters: the flux $(\Phi_{\mathrm{PL}})$ and spectral index $(\Gamma)$ of the power law, the flux $(\Phi_{\mathrm{SE}})$ and blackbody temperature ($kT$) of the soft excess, and the flux of the narrow Fe K$\alpha$ line $(\Phi_{\mathrm{K}\alpha})$. We initially allow $E_\mathrm{cut}$ to vary between 100--200 keV, based on the X-ray Spectral Energy Distrubution (SED) modeling results of NGC~7469 with Swift XRT+BAT in Prince et al. (in prep.). Their reported best-fit values using the \textsc{KYNSED} model \citep{Dovciak_2022} are $E_\mathrm{cut}=146\substack{+16\\ -6}$ keV for the spin $a^*=0$ case and $E_\mathrm{cut}=163\substack{+31\\ -48}$ keV for $a^*=1$, with a preference in $\chi^2_\nu$ for the $a^*=0$ model. Our spectra are insensitive to the cutoff energy, with errors overlapping with the edge of the allowed parameter range in 100\% of our epochs. 

We then compare the quality of fit using $E_\mathrm{cut}$ values of 146 keV and 163 keV, and using a power law model with no cutoff (\textsc{cutoffpl} and \textsc{powerlaw} in \textsc{XSPEC}, respectively). Since $E_\mathrm{cut}$ is fixed, the degrees of freedom for a given epoch are the same across the three models. This allows for a direct comparison of $\Sigma\chi^2$, the total best fit $\chi^2$ value summed across all 164 epochs for each model. With 16,124 total degrees of freedom, $\Sigma\chi^2=$16,363 for $E_\mathrm{cut}$=146 keV, $\Sigma\chi^2=$16,367 for $E_\mathrm{cut}$=163 keV, and $\Sigma\chi^2=$16,402 for a power law with no cutoff. Since we find a preference in $\chi^2$ for $E_\mathrm{cut}=146$ keV, we set this as the fixed value in our final model.

We measure the significance of the Fe K$\alpha$ line in our model using an F-test, comparing the $\chi^2$ for each epoch when fit with and without the narrow line. In our sample, 68 epochs  have a $p$-value~$<0.1$. This motivates our decision to include the narrow Gaussian in our final model. The final \textsc{XSPEC} model reads: \textsc{tbabs*(cflux*cutoffpl+cflux*blackbody+
cflux*gaussian)}.

\begin{deluxetable}{ccc}[t]
\tablewidth{0pt}
\tablecaption{\label{table:modelparams}Source Model Parameters}
\tablehead{
\colhead{Component} & \colhead{Parameter} & \colhead{Allowed Range}}
\startdata
\textsc{tbabs} & $N_\mathrm{H}$ ($10^{22}$cm$^{-2}$) &  0.0448 \tablenotemark{a} \\
\hline
\textsc{cflux} & $\log(\Phi_{\mathrm{PL}})$ (erg~cm$^{-2}$~s$^{-1}$) & $[-16,-8]$\\
\textsc{cutoffpl} & $\Gamma$ & $[1,3]$\\
& $E_\mathrm{cut}$ (keV) & $[100,200]$ \tablenotemark{b}\\
\hline
\textsc{cflux} & $\log(\Phi_{\mathrm{SE}})$ (erg~cm$^{-2}$~s$^{-1}$) & $[-16,-8]$\\
\textsc{blackbody} & $kT$ (keV) & $[0.08, 0.16]$ \tablenotemark{c}\\
\hline
\textsc{cflux} & $\log(\Phi_{\mathrm{K}\alpha})$ (erg~cm$^{-2}$~s$^{-1}$) & $[-16,-8]$ \tablenotemark{c}\\
\textsc{zgauss} & line energy (keV) &  6.4 \tablenotemark{a} \\
 & line width (keV) &  0 \tablenotemark{a}\\
 & redshift (z) &0.0163 \tablenotemark{a}\\
\hline
\enddata
\tablenotetext{a}{Always fixed}
\tablenotetext{b}{Fixed to $E_\mathrm{cut}=146$ keV (Prince et al. in prep.) in final model.}
\tablenotetext{c}{Fixed to mean best-fit values ($kT=0.12$ keV and \\$\log(\Phi_{\mathrm{K}\alpha})=-12.48$) in the final model.}
\end{deluxetable}

We define the spectra as sensitive to changes in a given spectral model parameter if the standard deviation of its best-fit value across all epochs ($\sigma_\mathrm{std}$) is twice its mean error at 68\% confidence ($\sigma_\mathrm{mean}$). The $1\sigma$ error must also  be constrained in $>90\%$ of observations, e.g. not fixed at the edge of the allowed range listed in Table~\ref{table:modelparams}.

In this manner, our spectra are sensitive to changes in $\Phi_{\mathrm{PL}}$, $\Gamma$,  and $\Phi_{\mathrm{SE}}$ (see Appendix~\ref{sec:swiftappendix}). These are left free in the final model. Our data do not show significant changes in $\Phi_{\mathrm{K}\alpha}$, since the errors are unconstrained for 34\% of observations and $\sigma_\mathrm{std}/\sigma_\mathrm{mean}=1.7$. This is reasonable since the narrow Fe K$\alpha$ line is faint. We fix $\Phi_{\mathrm{K}\alpha}$ to the mean best-fit value ($3.3 \times 10^{-13}$ erg~cm~s$^{-1}$). Changes in $kT$ of the soft excess are also below the significance threshold ($\sigma_\mathrm{std}/\sigma_\mathrm{mean}=1.5$), so we fix this parameter to its mean value of 0.12 keV. 

\begin{figure}[t]
  \centering
  \includegraphics[trim={1.2cm 1.2cm  0 0},clip,width=\columnwidth]{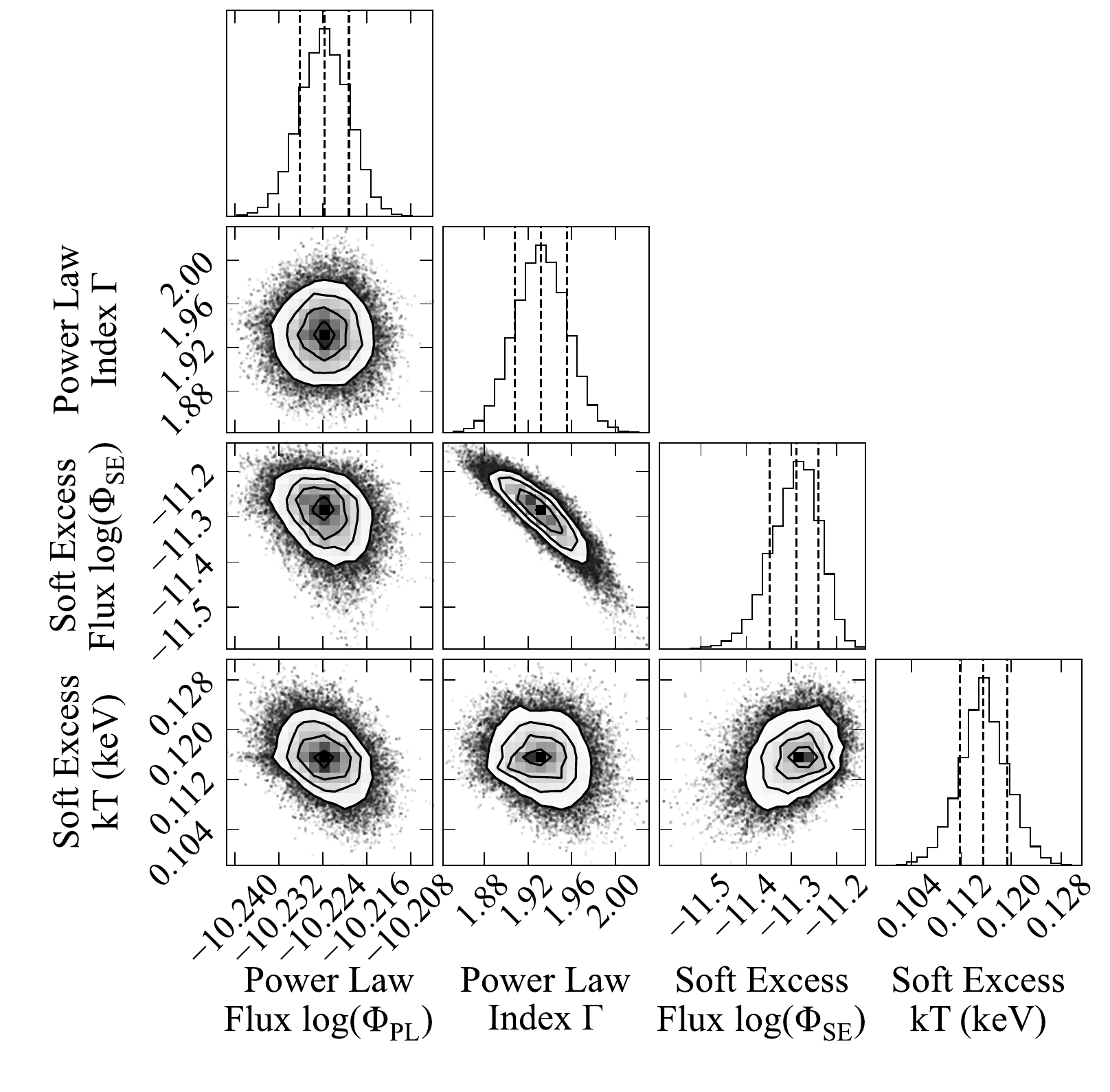}
  \caption{Corner plots of the free parameters in our spectral model, fit to an example NICER epoch on MJD 60104 (ObsID: 6714013101). We use the Markov Chain Monte Carlo fitting method in XSPEC to determine the probability distributions of each parameter, with a burn-in of 20,000 steps and 100,000 steps in total. A degeneracy is visible between the slope of the power law, denoted by index $\Gamma$, and the flux of the soft excess component from 0.3--10 keV ($\Phi_{\mathrm{SE}}$).}
  \label{fig:contourdegen}
\end{figure}

\begin{figure*}[t]
    \centering
    \includegraphics[width=0.9\textwidth]{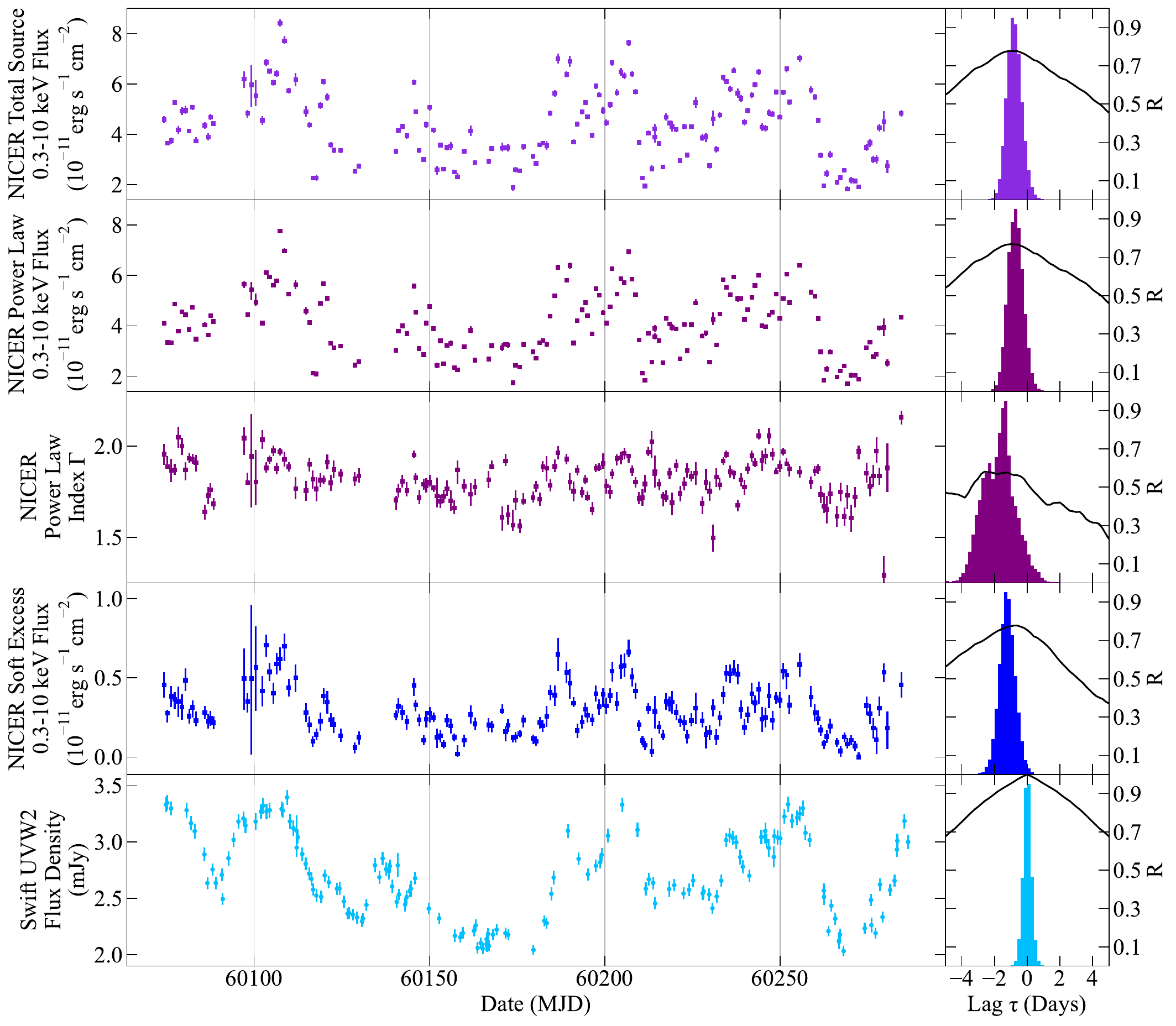} 
    \caption{\textbf{\textit{Left}}: Light curves of the best-fit flux values measured from 0.3--10 keV at each epoch of observation with NICER, using the model in Table~\ref{table:modelparams}. In descending order, this figure shows the total background-subtracted source flux (Panel 1, light purple), the power law flux ($\Phi_\mathrm{PL}$, Panel 2) and spectral index ($\Gamma$, Panel 3) in dark purple, and the soft excess flux ($\Phi_\mathrm{SE}$, Panel 4) in dark blue. The Swift UVW2 light curve is presented in light blue (Panel 5) as a flux density in units of mJy with central wavelength $\lambda=1928$ \AA. \textbf{\textit{Right}}: Interpolated cross-correlation functions measured with respect to the UVW2 reference band, calculated such that a negative lag indicates that the X-rays lead the UV. The probability distribution of the centroid lag values ($\tau_\mathrm{cent}$) are shown as histograms. The lower right panel shows the UVW2 auto-correlation function.}  
\label{fig:nobinCCF}
\end{figure*}

We find a degeneracy between $\Gamma$ and $\Phi_{\mathrm{SE}}$ (see Figure~\ref{fig:contourdegen}). Although these spectral components are thought to originate from physically distinct regions of emission, their spectral shapes below 2 keV mean they are degenerate in many AGN (e.g. \citealt{Partington_2024}) since a flatter power law model will require a stronger soft excess to achieve an equally good fit. However, the soft excess of NGC~7469 is bright enough that the $\Gamma$ and $\Phi_{\mathrm{SE}}$ are well-constrained in each spectrum. The correlation between the parameters across all epochs is also weak ($R=0.4$), so we leave both parameters free. The other parameters are well-constrained and do not show degeneracies. We present the best-fit spectral parameter values of our final model for each epoch in Figure~\ref{fig:nobinCCF} and Table~\ref{table:nobinningbestfitparams}. 

\begin{deluxetable*}{cccccc}[t]
\tablewidth{0pt}
\tablecaption{\label{table:nobinningbestfitparams}Spectral Analysis Results}
\tablehead{
\colhead{Obs. Date} &
\colhead{Total Source Flux}&
\colhead{Power Law} &
\colhead{Power Law} &
\colhead{Soft Excess} &
\colhead{Best fit}\\ 
\colhead{(MJD)} &
\colhead{$\log(\Phi_{\mathrm{PL}}+\Phi_{\mathrm{SE}}+\Phi_{\mathrm{K}\alpha})$} &
\colhead{Flux $\log(\Phi_{\mathrm{PL}})$} &
\colhead{Index $\Gamma$} &
\colhead{Flux $\log(\Phi_{\mathrm{SE}})$} &
\colhead{$\chi^2/\mathrm{DoF}$}\\
\colhead{} &
\colhead{(erg~cm$^{-2}$~s$^{-1}$)} &
\colhead{(erg~cm$^{-2}$~s$^{-1}$)} &
\colhead{} &
\colhead{(erg~cm$^{-2}$~s$^{-1}$)} &
\colhead{}
}
\startdata
60074.34 & $-10.34\substack{+0.01\\ -0.01}$ & $-10.388\substack{+0.007\\ -0.006}$ & $1.96\substack{+0.06\\ -0.06}$ & $-11.34\substack{+0.07\\ -0.09}$ & $139.24/125$ \\ 
60075.28 & $-10.44\substack{+0.01\\ -0.01}$ & $-10.477\substack{+0.007\\ -0.003}$ & $1.89\substack{+0.06\\ -0.01}$ & $-11.56\substack{+0.03\\ -0.11}$ & $122.31/110$ \\ 
60076.38 & $-10.43\substack{+0.01\\ -0.02}$ & $-10.478\substack{+0.007\\ -0.010}$ & $1.87\substack{+0.06\\ -0.06}$ & $-11.42\substack{+0.07\\ -0.08}$ & $126.30/116$ \\ 
60077.34 & $-10.28\substack{+0.01\\ -0.01}$ & $-10.313\substack{+0.005\\ -0.002}$ & $1.87\substack{+0.03\\ -0.03}$ & $-11.44\substack{+0.05\\ -0.08}$ & $101.86/90$ \\ 
60078.38 & $-10.38\substack{+0.02\\ -0.02}$ & $-10.421\substack{+0.008\\ -0.008}$ & $2.05\substack{+0.06\\ -0.06}$ & $-11.45\substack{+0.10\\ -0.14}$ & $86.40/69$ \\ 
... & ... & ... & ... & ... & ...  
\enddata
\tablecomments{Best-fit fluxes, spectral parameters, and the minimum $\chi^2$/Degrees of Freedom (DoF) for each epoch of observation. Error ranges correspond to the 68\% certainty level ($1\sigma$) for each fit parameter. The first five rows are shown, and the entire dataset is available in machine-readable format.}
\end{deluxetable*}

\subsection{Cross-Correlation Analysis}
\label{sec:crosscorrelation}
We test the correlation between the UVW2 and X-ray light curves using pyCCF \citep{Sun_2018}, an implementation of the Interpolated Cross Correlation Function \citep{Peterson_1998}. Each X-ray light curve is linearly interpolated and shifted by a lag $\tau$, and the Pearson correlation coefficient $R(\tau)$ is measured between the shifted X-ray light curve and the UVW2 reference band light curve. Here, a negative lag indicates that variability in the X-rays occurs before, or ``leads," variability in the UV. We test a range of lags from $-15$ to 15 days and construct the Interpolated Cross Correlation Function (ICCF) from the measured $R$ values for each $\tau$ value (see Figure~\ref{fig:contourdegen}). We also perform the same test using the optical \textit{g}-band light curve from our ground-based monitoring as the reference. The full \textit{g}-band light curves and ICCFs are shown in Appendix~\ref{sec:gbandappendix}.

We report the centroid lag $\tau_\mathrm{cent}$ for each band, which is calculated over the range of lags such that $R$ is within 80\% of the peak $R$ value of the entire ICCF. Uncertainties in $\tau$ are calculated using the Flux Randomization (FR) and Random Subset Selection (RSS) methods, with 25,000 steps in the Monte Carlo algorithm used for RSS. Our $1\sigma$ uncertainties for $\tau_\mathrm{cent}$ reported in Tables~\ref{table:UVcentlags} and~\ref{table:gcentlags} are drawn from the probability distribution of the FR/RSS values, using the central 68\% of values around the reported median $\tau_\mathrm{cent}$. 

\begin{deluxetable*}{ccccc}[t]
\tablewidth{0pt}
\tablecaption{\label{table:UVcentlags}X-ray/UV Correlation and Lags}
\tablehead{
\colhead{} &
\multicolumn{2}{c}{No Smoothing}  &  \multicolumn{2}{c}{20 Day Boxcar Smoothing} \\
\colhead{Parameter} & 
\colhead{$R_\mathrm{cent}$} & 
\colhead{$\tau_\mathrm{cent}$ (days)} &
\colhead{$R_\mathrm{cent}$} & 
\colhead{$\tau_\mathrm{cents}$ (days)}
}
\startdata
Total Flux& $0.78$&$-0.8\substack{+0.5\\ -0.4}$& $0.94$&$-2.2\substack{+0.6\\ -0.2}$\\ 
Power Law Flux& $0.77$&$-0.7\pm0.5$& $0.94$&$-2.2\substack{+0.6\\ -0.2}$\\ 
Power Law Index& $0.57$&$-1.7\substack{+0.9\\ -1.1}$& $0.91$&$-1.0\substack{+0.8\\ -0.6}$\\ 
Soft Excess Flux& $0.77$&$-1.2\pm0.5$& $0.91$&$-2.2\substack{+0.6\\ -0.4}$\\ 
\hline
\enddata
\tablecomments{Centroid lag values for NICER spectral parameters and fluxes from 0.3--10 keV. The Swift UVW2 band is used as a reference, and negative lags indicate that the X-ray variability leads the UV.}
\end{deluxetable*}

\begin{deluxetable}{ccc}[t]
\tablewidth{0pt}
\tablecaption{\label{table:gcentlags}X-ray/Optical Correlation and Lags}
\tablehead{
\colhead{Parameter} & 
\colhead{$R_\mathrm{cent}$} & 
\colhead{$\tau_\mathrm{cent}$ (days)}
}
\startdata
Total Flux& $0.72$&$-1.3\substack{+0.4\\ -0.3}$\\ 
Power Law Flux& $0.71$&$-1.2\pm0.4$\\ 
Power Law Index& $0.54$&$-2.2\pm0.8$\\ 
Soft Excess Flux& $0.73$&$-1.7\pm0.4$\\ 
UVW2 Flux Density& $0.92$&$-0.4\pm0.8$\\ 
\hline
\enddata
\tablecomments{Centroid lag values for NICER spectral parameters and fluxes from 0.3--10 keV. The optical \textit{g}-band (central wavelength $\lambda=4770$ \AA) is used as a reference, and negative lags indicate that the X-ray and UV variability leads the optical.}
\end{deluxetable}

The correlation between the flux of each X-ray model component and the UVW2 band is high ($R_\mathrm{cent}=0.78$ for the power law and $R_\mathrm{cent}=0.77$ for the soft excess), with lags near zero days. The power law index is modestly correlated with the UVW2 ($R=0.57$), likely as a consequence of the softer-when-brighter behavior of Comptonization in the corona. The correlation between the X-ray power law flux and the \textit{g}-band is weaker ($R_\mathrm{cent}=0.71$) with a longer lag of $\tau_\mathrm{cent}=-1.2\pm0.4$ days. This is expected if the optical emission originates at larger radii than UV source, relative to the black hole.

\begin{figure}[t]
  \centering
  \includegraphics[width=0.9\columnwidth]{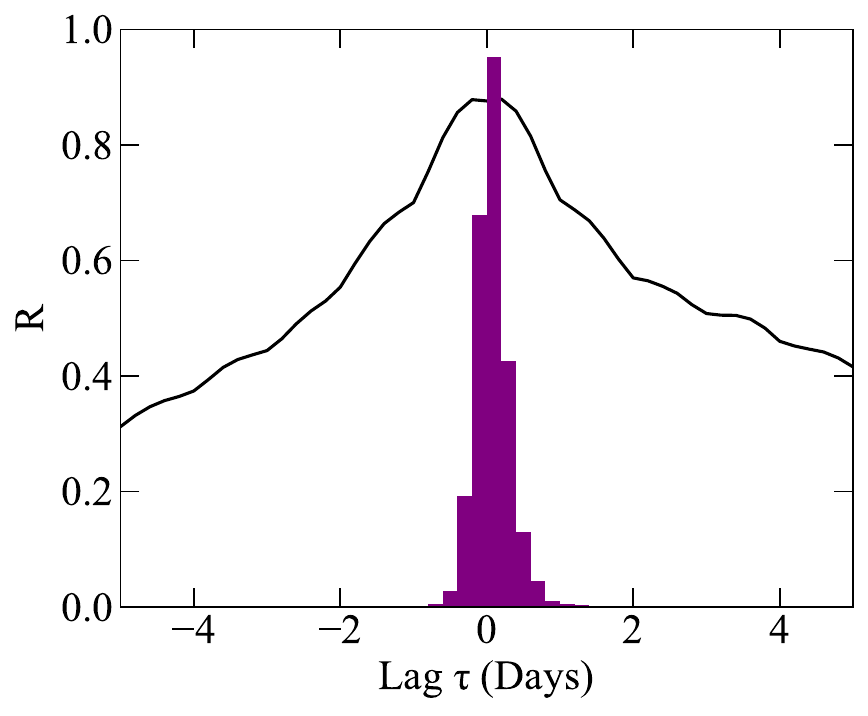} 
  \caption{ICCF results between the X-ray power law and soft excess using the light curves in Fig~\ref{fig:nobinCCF}. The soft excess is used as the reference light curve. The centroid lag is $\tau=0.1 \pm 0.2$ days, with $R=0.88$.}
  \label{fig:PL_SErefband_ccf}
\end{figure}

We also test for a relationship between the X-ray fluxes of the power law and the soft excess to probe the nature of the latter emission source. Using the soft excess as the reference band, we measure a steeply peaked ICCF near a lag of zero days, with $R_\mathrm{cent}=0.88$ (Fig.~\ref{fig:PL_SErefband_ccf}). Notably, the power law emission is more strongly correlated with the soft excess than the UVW2. The implications of these interband lags and correlations are discussed in Section~\ref{sec:Discussion}.

\begin{figure}[t]
  \centering
  \includegraphics[width=\columnwidth]{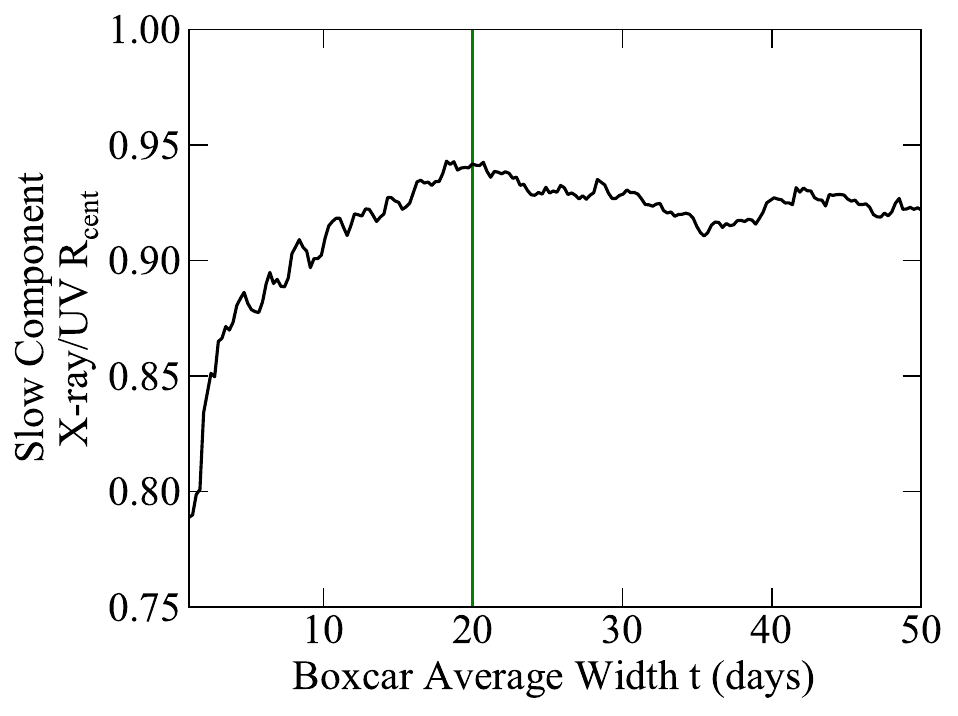}
  \caption{Correlation $R$ between the boxcar-smoothed X-ray power law and UVW2 light curves, as a function of the boxcar smoothing timescale $t$ from 1--50 days. The maximum $R_\mathrm{cent}\simeq0.94$ occurs from $t=18$--20 days (green line).}
  \label{fig:UVX-rayCCF_Run103_boxcarrange}
\end{figure}

We compare slow variability in the X-rays and the UV by smoothing the light curves over a range of common timescales from $t=1$--50 days. For each timescale ($t$), we perform a rolling boxcar average on the light curve by replacing the flux at every epoch with the mean flux over all epochs within $\pm t/2$ days. To ensure an even smoothing timescale throughout the light curve, we exclude all epochs within $\pm t/2$ days of the start and end points of observation from the final light curve. We measure the ICCF across a lag ranging from $\tau=\pm 50$ days for each smoothing timescale and present the corresponding $R_\mathrm{cent}$ values in Figure~\ref{fig:UVX-rayCCF_Run103_boxcarrange}. The maximum  X-ray/UV correlation of $R_\mathrm{cent}=0.94$ is measured when the light curves are smoothed by a width of 18--20 days.

\begin{figure*}[t]
    \centering
    \includegraphics[width=0.9\textwidth]{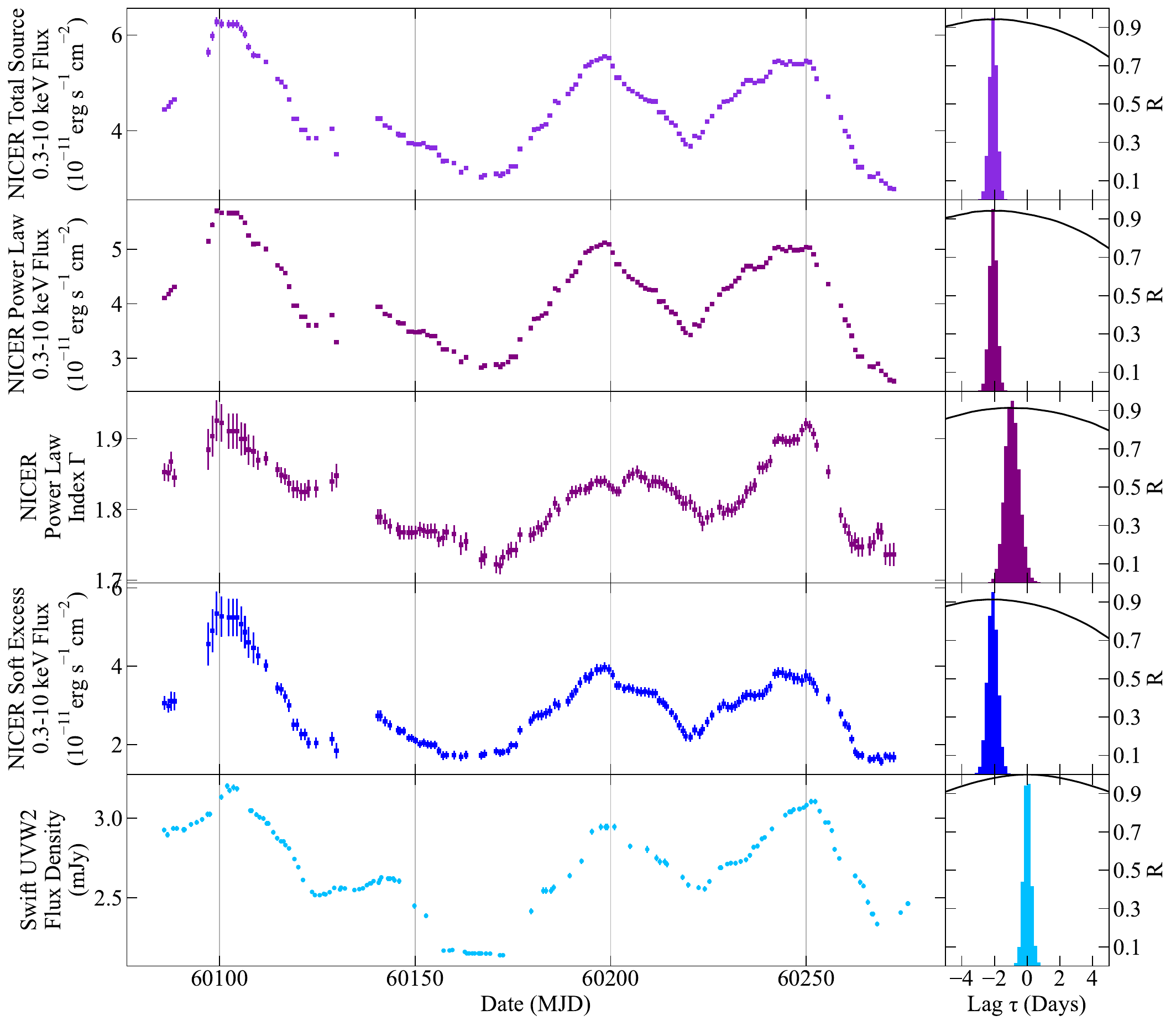} 
    \caption{The same as Figure~\ref{fig:nobinCCF}, but for X-ray and UV light curves which have been smoothed by performing a rolling boxcar average with a total width of $t=20$ days. This is done to isolate slow variability in the AGN. The variability amplitude is smaller here than it is in the raw light curve, and the ICCFs are significantly smoother and flatter. The maximum $R_\mathrm{cent}$ between the smoothed UVW2 light curve and each of the smoothed X-ray components is higher than $R_\mathrm{cent}$ of the raw light curves, as listed in Table~\ref{table:UVcentlags}.}  
\label{fig:slowbinCCF}
\end{figure*}

The boxcar-smoothed light curves that show variations on timescales longer than 20 days are presented in Figure~\ref{fig:slowbinCCF}. The negative lags (such that the X-rays lead the UV) are longer than those from the raw light curves (see Table~\ref{table:UVcentlags}). However, the ICCFs are also extremely flat as a result of the averaging method, and as such these measurements are not likely to be a reliable indicator of the physical scale of the disk.

\subsection{Energy and Variability of the Disk and Corona}
\label{sec:egyvaranalysis}
To test if the X-ray corona can drive the variability in the accretion disk on timescales of a few days, as predicted by the X-ray reprocessing scenario, we quantify the energies of the variable emission components in each region. Prince et al. (in prep.) estimates the total luminosity of the disk ($L_{\mathrm{Disk}}$) to be 21\% of the Eddington luminosity ($L_{\mathrm{Edd}}=1.13\times 10^{45}$ erg~s$^{-1}$), i.e. $L_{\mathrm{Disk}}=2.38 \times 10^{44}$ erg~s$^{-1}$, and the source to be located at a luminosity distance  ($D_\mathrm{L}$) of 70.5 Mpc. Their method fits the UV and optical inter-band lags and mean SED to a ``bowl-shaped" disk model with a concave vertical thickness profile. These values assume $M_\mathrm{BH}=9 \times 10^6 M_\odot$ \citep{Bentz15} and $z=0.0163$ \citep{Springob_2005}. 

During the Prince et al. (in prep.) campaign from MJD 59345--59602, the mean Swift UVW2 flux density is $\overline{\phi}_{\nu,\mathrm{UVW2}}=2.70\pm0.01$ mJy. From MJD 60074–-60284, we measure $\overline{\phi}_{\nu,\mathrm{UVW2}}=2.703\pm0.005$ mJy. Since the central frequency of the UVW2 band is $1.55\times 10^{15}$ Hz ($\lambda=1928$~\AA ), the mean flux is $\overline{\Phi}_{\mathrm{UVW2}}=4.20\times 10^{-11}$ erg~cm$^{-2}$~s$^{-1}$. The mean flux is consistent between the campaigns, indicating that the accretion rate and disk luminosity of \ngc\ are stable. As such, we can use $L_{\mathrm{Disk}}$ and $D_\mathrm{L}$ from Prince et al. (in prep.) to estimate the total disk flux, $\Phi_{\mathrm{Disk}}=4.01\times 10^{-10}$ erg~cm$^{-2}$~s$^{-1}$, since $\Phi_{\mathrm{Disk}}=L_{\mathrm{Disk}}/(4\pi D^2_\mathrm{L}$). 

Figure~7 of Prince et al. (in prep.) shows the modeled SED of the X-ray corona with \textsc{KYNSED}, using averaged Swift XRT+BAT spectra from 0.3--200 keV. Their best-fit SED for a spin $a^*=0$ black hole includes a corona with low- and high-energy cutoffs of 0.03 keV and 146 keV, and a power law index $\Gamma=1.80\substack{+0.02\\ -0.03}$. During our campaign with NICER, the mean flux of the unobscured power law from 0.3--10 keV is $\overline{\Phi}_{\mathrm{PL}}=4.04\times 10^{-11}$ erg~cm$^{-2}$~s$^{-1}$, and the mean $\Gamma=1.82$. To estimate the mean flux of the total X-ray corona with NICER, we extrapolate the mean spectrum out to 0.03--200 keV using the \textsc{energies} command in XSPEC. We obtain $\overline{\Phi}_{\mathrm{Corona}}=9.75\times 10^{-11}$ erg~cm$^{-2}$~s$^{-1}$ with the XSPEC \textsc{flux} command. We also measure the flux of the soft excess $\overline{\Phi}_{\mathrm{SE}}=4.09\times 10^{-12}$ erg~cm$^{-2}$~s$^{-1}$ from 0.03--200 keV, based on a mean flux from 0.3--10 keV of $2.93\times 10^{-12}$ erg~cm$^{-2}$~s$^{-1}$ and blackbody temperature $kT=0.12$ keV. 

We measure the amount of variability in a given light curve using the mean fractional variation, 
\begin{equation}
    F_\mathrm{var}=\sqrt{\frac{S^2-\overline{\sigma^2_{\mathrm{err}}}}{\overline{\Phi}^2}}
\end{equation}
(\citealt{Rodriguez_1997}, \citealt{vaughan03}). Here, $S^2$ is the variance of the light curve, $\overline{\sigma^2_{\mathrm{err}}}$ is the mean square error, and ${\overline{\Phi}}$ is the mean flux across all epochs. We use $F_\mathrm{var}$ of the power law (31.1\%) and the soft excess (46.3\%) flux light curves from 0.3--10 keV as proxies for their total variability in the extended energy range. 

The corresponding change in flux associated with $F_\mathrm{var}$, relative to the mean of the light curve, is 
\begin{equation}
    \Delta(\Phi)={\overline{\Phi}}\times F_\mathrm{var}.
\end{equation}
In the X-ray corona, $\Delta(\Phi_{\mathrm{Corona}})=3.03\times 10^{-11}$ erg~cm$^{-2}$~s$^{-1}$, and $\Delta(\Phi_{\mathrm{SE}})=1.89\times 10^{-12}$ erg~cm$^{-2}$~s$^{-1}$ in the soft excess.

The UVW2 band has $F_\mathrm{var}=13.6\%$, corresponding to a change of $\Delta(\Phi_{\mathrm{UVW2}})=5.73\times 10^{-12}$ erg~cm$^{-2}$~s$^{-1}$. For the UVW2 variability to be powered by thermal reprocessing of X-rays, the variable X-ray flux must exceed it, i.e. $\Delta(\Phi_{\mathrm{Corona}}) \geq \Delta(\Phi_{\mathrm{UVW2}})$. This is clearly feasible, with the X-rays exceeding the UV by $~\sim5\times$. 
Considering only the slow variations using a boxcar smoothing timescale of 20 days (see Sec~\ref{sec:crosscorrelation}), $F_\mathrm{var,UVW2,slow}=10.6\%$ and $F_\mathrm{var,Corona,slow}=19.1\%$. The slow variation of the X-ray flux is then $~\sim4\times$ greater than the UVW2, suggesting that X-ray reprocessing can also drive slow changes in the UV light curve on timescales of weeks.

For the X-rays to power fast variability in the entire disk, it is required that $\Delta(\Phi_{\mathrm{Corona}}) \geq \Delta(\Phi_{\mathrm{Disk}})$. We can restructure Equation (2) to obtain the maximum mean fractional variation of the entire disk 
\begin{equation}
F_\mathrm{var}\leq\frac{\Delta(\Phi_{\mathrm{Corona}})}{\overline{\Phi}_{\mathrm{Disk}}},
\end{equation}
or $7.6$\%. This threshold requires a much smaller variability amplitude in the optical light curve than the UV. Our \textit{g} band measurements support this scenario, with $F_\mathrm{var}=4.2\%$.

We estimate the relative UV flux contributions of both the AGN and nuclear starburst regions of NGC 7469 using an archival HST image taken on MJD 51357 with the F218W filter (central wavelength $\lambda=2228$ \AA) and the WFPC2 instrument. Both the AGN and the starburst are constrained within the $10\arcsec$ aperture radius chosen for our Swift UVOT photometry (see Section~\ref{sec:Data}). The AGN emission contributes $65 \pm 14\%$ of the observed flux and is constrained within a radius of $0.5\arcsec$. The remaining flux within the aperture originates from the starburst. The relative contribution of the starburst is expected to decline at shorter wavelengths, such as the Swift UVW2 filter at 1928 \AA\ \citep{Mehdipour_2018}. Since the constant component of the UVW2 light curve in Figure~\ref{fig:nobinCCF} originates partially from the starburst, our results serve as a lower limit on the possible fractional variability in the AGN disk.

\section{Discussion}
\label{sec:Discussion}
\subsection{The X-ray Reprocessing Scenario}
\label{sec:xrayreprodiscuss}
We find a strong correlation between the X-ray power law and the UVW2 continuum ($R=0.76$), with the X-rays leading the UV. This suggests that the variability is linked through reprocessing of an outbound X-ray reverberation signal. Our analysis shows that the X-rays can drive variability on timescales of days in the UVW2 band through thermal reprocessing, since the X-ray corona has sufficient energy and variability to explain the observed correlation. The X-rays are more strongly correlated with the UVW2 band ($R=0.76$) than the \textit{g}-band ($R=0.71$), suggesting that the reprocessed variability signal is smoothed at large radii. This trend was also observed during previous campaigns on the source (e.g., Prince et al. in prep.). 

The X-rays may have enough power to drive changes in the entire disk since the fractional variability decreases at larger disk radii, where emission peaks in the optical. This is expected since the flux per unit area arriving at a given location falls off following $R^{-2}$. Additionally, since the X-ray corona lies above the plane of the disk at a height of a several gravitational radii ($R_\mathrm{G}$, \citealt{kara16}), a region of the disk with width $\Delta R$ will subtend a smaller solid angle at larger $R$ for $R\gtrsim100 R_\mathrm{G}$. This means that the most dramatic variability is expected to occur in the inner disk. This trend is seen in the UV/Optical variability measurements for \ngc\ in Section~\ref{sec:egyvaranalysis}. Similarly, Prince et al. (in prep.) showed that the ratio between the root mean square variability amplitude and the mean flux of each light curve drops steeply with increasing wavelength. Our study and Prince et al. (in prep.) also found that the fractional amplitude of the observed UV/optical variability may be affected in part by galaxy contamination.

Averaging the light curve over timescales of 20 days improves the X-ray/UV correlation to $R=0.94$. This suggests that a common process governs slow variability in both sources, with additional fast variability in the X-rays not seen in the UV. In an analysis of Fairall~9, \cite{Hagen_2024} suggest that the slow variability originates from instabilities in the accretion disk which modulate the flux of the X-ray corona. This would cause the X-rays to lag behind the UV on long smoothing timescales. In a re-analysis of the RXTE campaign on \ngc\ presented in \cite{Nandra_2000}, \cite{Pahari_2020} find that the UV leads the X-rays on timescales longer than 5 days, while the remaining fast variability propagates in the direction of reverberation.

However, our results in Figure~\ref{fig:slowbinCCF} show a ICCF with negative lags, in the opposite direction expected for propagating disk fluctuations. This suggests that the variability on short and long timescales may be produced by reverberation, directly contradicting the results of \cite{Pahari_2020}. The diminished fast variability in the UV could then arise from the inherent blurring of the signal as multiple disk radii contribute to the UVW2 flux, reducing the impact of rapid and small changes in the flux incident on the disk. We do note that the ICCF is extremely flat as a consequence of our boxcar averaging method, so our recovered lag may not be reliable. In the future, a Fourier analysis of the raw light curves would more reliably measure a lag in the slow signal and may be able to distinguish between the described scenarios. 

\subsection{Geometry of the Accretion Disk}
\label{sec:geomdiscuss}
Since the variability in the accretion disk is consistent with reprocessed emission from the X-ray corona, we can estimate the distance between these regions using the measured X-ray/UV lag of $\tau=-0.7 \pm 0.5$ days (Table~\ref{table:UVcentlags}). We convert $\tau$ to a light travel distance of $1364 \pm 975 R_\mathrm{G}$, assuming $M_\mathrm{BH}=9 \times 10^6 M_\odot$ \citep{Bentz15}. We then compare this with the theoretical prediction for the disk radius where emission peaks in the UVW2 band ($\lambda=1928$ \AA{}) using Equation (12) of \cite{fausnaugh16}:

\begin{equation} 
    R=\left(X\frac{k\lambda}{hc}\right)^{4/3}\left[\left(\frac{GM}{8\pi\sigma}\right) \left(\frac{L_\mathrm{Edd}}{\eta c^2}\right)(3+\kappa)\dot{m}_\mathrm{Edd}\right]^{1/3}. 
\end{equation}

The multiplicative factor $X$ converts between the disk temperature and the emission wavelength $\lambda$. This accounts for the radially extended region of the disk which produces some emission at this wavelength, with the majority originating the radius $R$ whose blackbody flux peaks at $\lambda$. We compare results from two derivations of $X$, which account for the mean radius of peak emission using Planck's law ($X=2.49$, \citealt{fausnaugh16}) and the effect of temperature perturbations from a variable heating source ($X=5.04$, \citealt{tie_2018}). We assume a typical efficiency of $\eta=0.1$ for the conversion of rest mass into energy and comparable amounts of heating from internal viscous interactions and the external X-ray source ($\kappa=1$). 

We first use the Eddington ratio ($\dot{m}_\mathrm{Edd}=L_\mathrm{Bol}/L_\mathrm{Edd}$) of 0.16 from the spin $a^*=0$ \textsc{KYNSED} model in Prince et al. (in prep.) The resulting predictions for $R_\mathrm{UVW2}$ are 0.08 light days (155 $R_\mathrm{G}$) for $X=2.49$ and 0.20 light days (390 $R_\mathrm{G}$) for $X=5.04$. The value of $\dot{m}_\mathrm{Edd}$ measured in Prince et al. (in prep.) is highly dependent on the SED model. For instance, $\dot{m}_\mathrm{Edd}=0.13$ in the spin $a^*=1$ \textsc{KYNSED} model, and the luminosity of the disk alone in the Bowl model is $0.21L_\mathrm{Edd}$. Combining the flux of the X-ray corona measured with NICER in Section~\ref{sec:egyvaranalysis} and the disk flux from the Bowl model, we obtain $\dot{m}_\mathrm{Edd}=0.26$. For $X=5.04$, the theoretical value for $R_\mathrm{UVW2}$ is then 0.24 light days (468 $R_\mathrm{G}$).

We perform the same test using the X-ray/optical lag of $-1.2\pm0.4$ days ($2338\pm780R_\mathrm{G}$), using the \textit{g}-band as the reference ($\lambda=4770$ \AA{}). The theoretical location of $R_\mathrm{\textit{g}-band}$ obtained using $X=5.04$ and $\dot{m}_\mathrm{Edd}=0.26$ most closely matches the observed value, yielding $R_\mathrm{\textit{g}-band}=0.79$ light days (1540 $R_\mathrm{G}$).

Given their large uncertainties, our measured X-ray/UV and X-ray/optical lags are broadly consistent with the theoretical locations of the peak UVW2 and \textit{g}-band emission. However, we are not able to observationally distinguish between values of $X$ or $\dot{m}_\mathrm{Edd}$. This arises from the limited precision of our lag measurements, given that our cadence of one day exceeds the observed lag of 0.7 days.

\subsection{The Soft Excess Source}
Many other AGN display weakly correlated X-ray and UV light curves, and an additional optically thick region located in between the central X-ray corona and the disk is often invoked to explain this behavior. In heavily obscured sources, this may be an inner disk wind (e.g. Mrk~817, \citealt{partington23}). In ``bare" AGN, this may be caused by a vertically extended layer of electrons above the inner disk with lower temperatures and a higher optical depth than those in the central X-ray corona. Sometimes referred to as the ``warm corona," this region is expected to emit flux in the soft X-rays through Comptonization, contributing to the soft excess. 

For instance, using energy-binned light curves of Mrk~509 with XMM-Newton, \cite{mehdipour11} found a strong correlation between the soft X-rays and UV which decreased at higher X-ray energies. This suggested the presence of such a region. During campaigns on NGC~5548 \citep{Gardner2017} and NGC~4151 \citep{edelson2017} using Swift, this optically thick layer was invoked to explain the length of UV/Optical lags and the fast X-ray variability signal which was not present in the disk.  In Fairall~9, a time-domain spectral modeling campaign with NICER revealed that the UVW2 light curve was more strongly correlated with the soft excess than the power law continuum, suggesting the presence of a warm corona \citep{Partington_2024}. 

In contrast our analysis of \ngc\ shows strongly correlated ($R\simeq0.8$) fast X-ray/UV variability, providing no evidence for an intermediate reprocessor. We also measure a strong correlation ($R\simeq0.9$) between the X-ray power law and soft excess and a lag consistent with zero ($\tau=0.1 \pm 0.2$) days  (Fig.~\ref{fig:PL_SErefband_ccf}), suggesting that the soft excess originates from reflection off of the inner disk. The lack of evidence for a warm corona in \ngc\ suggests that the inner accretion flows of AGN do not follow a uniform prescription, since some are consistent with the standard reverberation picture while others require additional physical components. 

The power law/soft excess lag corresponds to an upper limit on the light travel distance between the emission sources of 194 $R_\mathrm{G}$ if the power law leads the soft excess (i.e. the lag is negative). We are limited by our daily observation cadence, which prevents a more precise measurement of this possible reverberation signal. A reverberation study by \cite{kara16} uses an XMM-Newton observation with an exposure length of 163 ks to detect short lags at 5--7 keV, attributed to the relativistically broadened Fe K$\alpha$ reflection line from the inner accretion disk. The measured lag of 1848 ± 1451 s (0.005--0.038 days) can be interpreted as a lower limit on the light travel distance (9--75 $R_\mathrm{G}$), since it is not corrected for dilution by the continuum flux. While our NICER spectra are not sensitive to the presence of the broad Fe K$\alpha$ line, our lag window of $1\sigma$ between the power law and soft excess sources overlaps with the region measured in \cite{kara16}. This is consistent with a scenario in which the soft excess is dominated by the blurred reflection component.

In Section~\ref{sec:egyvaranalysis}, we found that the variable flux of the X-ray power law is $\sim16\times$ greater than that of the soft excess. This suggests that the X-ray corona has ample energy to power the soft excess variability through reflection. It is possible that the reflected emission below 0.3 keV is greater than the amount predicted by our blackbody model, since it does not contain physical parameters of the source which could be extrapolated to lower energies. Future improvements to the analysis using an instrument with better spectral sensitivity, particularly around the Fe K$\alpha$ line, such as the proposed STROBE-X, would enable tests of physical reflection models such as \textsc{Relxill} \citep{garcia16} at a daily cadence.

\section{Summary and Conclusions} 
\label{sec:Conclusion}
We model the X-ray power law and soft excess components of \ngc\ using NICER spectra taken daily from 10 May 2023 to 06 Dec 2023, in combination with contemporaneous Swift UVW2 band monitoring. Our results show that:

\begin{itemize}
    \item The X-ray power law index $\Gamma$ is modestly correlated ($R=0.57$) with the UVW2 light curve, following the ``softer-when-brighter" AGN pattern, with a flat ICCF near the peak (Figure~\ref{fig:nobinCCF}). This suggests that the X-ray emission arises due to the Comptonization of seed photons from the inner disk in a hot electron corona.
    \item Variability in the X-ray corona is strongly correlated with the UVW2 continuum, and the X-rays lead by less than a day. We analyze the total flux and fractional variability amplitude of both sources (Section~\ref{sec:egyvaranalysis}), demonstrating that the X-rays have sufficient energy to drive the UVW2 light curve through thermal reprocessing. 
    \item Slow variability on timescales of $\sim20$ days in the X-ray power law and the UVW2 band are correlated with $R=0.94$ (Figure~\ref{fig:slowbinCCF}). Given the large amount of available energy in the X-rays, these fluctuations do not necessarily have to originate from inwardly propagating fluctuations in the disk. This trend may instead arise from the intrinsic smoothing of the UV light curve features expected from X-ray reprocessing. However, given the extremely flat ICCF in our smoothed light curve, our lag measurements cannot sufficiently distinguish between these scenarios. 
    \item The ICCF between the X-ray fluxes of the power law and the soft excess peaks at $R=0.88$. The lag is consistent with zero days (Figure~\ref{fig:PL_SErefband_ccf}). This suggests that the soft excess originates from inner disk reflection. We find no evidence for a vertically extended layer of warm electrons above the inner disk (a so-called ``warm corona"), unlike the similar experiment with NICER performed on Fairall~9 by \cite{Partington_2024}.
\end{itemize}

\begin{acknowledgments}
ERP and EMC gratefully acknowledge funding for this analysis through NASA grant 80NSSC23K1043. ERP acknowledges support by the European Union (ERC, MMMonsters, 101117624).

\end{acknowledgments}
\facilities{NICER, Swift}
\software{Astropy \citep{Astropy_2013},
          Matplotlib \citep{Hunter_2007}, SciPy \citep{Scipy_2020}, PyCCF (\citealt{Peterson_1998}, \citealt{Sun_2018}), HEASOFT \citep{Blackburn_1995}, FTOOLS \citep{FTOOLS_2014}, ftgrouppha \citep{kaastra16}, Swift XRT Product Builder (\citealt{evans07}, \citealt{evans09}),  XSPEC \citep{arnaud96}, \textsc{tbabs} \citep{Wilms_2000}} 

\newpage
\appendix
\twocolumngrid
\section{Instrumental Sensitivity to Spectral Parameters}
\label{sec:swiftappendix}

We test if Swift is also sensitive to the X-ray soft excess in NGC~7469 using 0.3--10 keV spectra generated by the XRT Products Builder \citep{evans09} for each epoch between MJD 60074--60284. We use the same model as for the NICER spectra in Table~\ref{table:modelparams}, first fitting each epoch with the following free parameters: the power law flux $(\Phi_{\mathrm{PL}})$ and spectral index $(\Gamma)$, the blackbody soft excess flux $(\Phi_{\mathrm{SE}})$ and temperature ($kT$), and the narrow Fe K$\alpha$ line flux $(\Phi_{\mathrm{K}\alpha})$. The best-fit parameters for each epoch are shown in Table~\ref{table:swiftparam}.

To measure the instrumental sensitivity to changes in each parameter across the duration of the campaign, we compare the mean error corresponding to the 68\% confidence level ($\sigma_\mathrm{mean}$) and the standard deviation ($\sigma_\mathrm{std}$) across all epochs. For a spectrum to be considered sensitive to a given model parameter, we require the criteria discussed in Section~\ref{sec:specparam}: $\sigma_\mathrm{std}/\sigma_\mathrm{mean}>2$, with constrained errors (i.e. not pegged at the edge of the allowed range in Table~\ref{table:modelparams}) for at least 90\% of observations. 

We compare $\sigma_\mathrm{std}/\sigma_\mathrm{mean}$ in Table~\ref{table:sigmacomp} and the error constraint rate in Table~\ref{table:errorconstraincomp} between Swift and NICER. Swift measurements of the soft excess blackbody temperature ($kT$) nominally pass our variability threshold, however, 68\% of observations have errors pegged at the edge of the allowed [0.05,0.2] keV range for $kT$. This variability represents jitter within the extremes of the parameter range and is not a physically meaningful result. 

The low count rate sensitivity of Swift introduces large uncertainty in the flux of the soft excess, preventing the use of Swift spectra for reverberation mapping experiments on the separate contributions of individually modeled X-ray emission components. This means that our investigation can only be done with NICER.  

\begin{deluxetable*}{cccccccc}[t]
\tablewidth{0pt}
\tablecaption{\label{table:swiftparam}Swift XRT 0.3--10 keV light curve and best-fit spectral parameters using the model in Table~\ref{table:modelparams}.}
\tablehead{
\colhead{Obs. Date} &
\colhead{Count Rate} &
\colhead{Power Law} &
\colhead{Power Law} &
\colhead{Soft Excess} &
\colhead{Soft Excess} &
\colhead{Fe K$\alpha$} &
\colhead{Best fit}\\ 
\colhead{(MJD)} &
\colhead{(counts~s$^{-1}$)} &
\colhead{Flux $\log(\Phi_{\mathrm{PL}})$} &
\colhead{Index $\Gamma$} &
\colhead{Flux $\log(\Phi_{\mathrm{SE}})$} &
\colhead{$kT$} &
\colhead{Flux $\log(\Phi_{\mathrm{K}\alpha})$} & 
\colhead{C-stat/DoF}\\
\colhead{} &
\colhead{(c s$^{-1}$)} &
\colhead{(erg~cm$^{-2}$~s$^{-1}$)} &
\colhead{} &
\colhead{(erg~cm$^{-2}$~s$^{-1}$)} &
\colhead{(keV)} &
\colhead{(erg~cm$^{-2}$~s$^{-1}$)} &
\colhead{}
}
\startdata
60074.85 & $1.0\pm0.1$ & $-10.46\substack{+0.05\\ -0.04}$ & $1.4\substack{+0.3\\ -0.3}$ & $-11.2\substack{+0.2\\ -0.6}$ & $0.19\substack{+0.01\\ -0.04}$ & $-16.0\substack{+4.1\\ -0.0}$ & $43.90/207$ \\ 
60075.18 & $1.0\pm0.1$ & $-10.41\substack{+0.03\\ -0.03}$ & $1.9\substack{+0.1\\ -0.1}$ & $-11.6\substack{+0.3\\ -1.5}$ & $0.05\substack{+0.05\\ -0.00}$ & $-12.4\substack{+0.7\\ -3.6}$ & $24.08/156$ \\ 
60076.30 & $0.9\pm0.1$ & $-10.46\substack{+0.04\\ -0.04}$ & $1.7\substack{+0.2\\ -0.2}$ & $-11.5\substack{+0.3\\ -1.0}$ & $0.12\substack{+0.05\\ -0.05}$ & $-12.8\substack{+1.0\\ -3.2}$ & $45.22/301$ \\ 
60077.21 & $1.8\pm0.5$ & $-10.19\substack{+0.14\\ -0.15}$ & $1.6\substack{+0.5\\ -0.6}$ & $-11.1\substack{+0.4\\ -4.9}$ & $0.20\substack{+0.00\\ -0.15}$ & $-11.0\substack{+0.5\\ -5.0}$ & $13.59/298$ \\ 
60080.73 & $1.4\pm0.1$ & $-10.31\substack{+0.03\\ -0.03}$ & $1.8\substack{+0.1\\ -0.1}$ & $-11.9\substack{+0.4\\ -4.1}$ & $0.05\substack{+0.15\\ -0.00}$ & $-11.8\substack{+0.4\\ -4.2}$ & $42.02/260$ \\ 
... & ... & ... & ... & ... & ... & ... & ...
\enddata
\tablecomments{Count rates and spectra are generated using the XRT Product Builder (\citealt{evans07}, \citealt{evans09}). Listed errors correspond to the $1\sigma$ uncertainty (68\% confidence) level. The first five epochs are shown, and the full table is available in machine-readable format.}
\end{deluxetable*}

\begin{deluxetable*}{cccccccc}[t]
\tablewidth{0pt}
\tablecaption{\label{table:sigmacomp}Sample Variability and Instrument Sensitivity}
\tablehead{
\colhead{Instrument} &
\colhead{Number of} &
\colhead{Mean Source} &
\colhead{Power Law} &
\colhead{Power Law} &
\colhead{Soft Excess} &
\colhead{Soft Excess} &
\colhead{Fe K$\alpha$}\\ 
\colhead{} &
\colhead{Epochs} &
\colhead{Count Rate} &
\colhead{Flux $\log(\Phi_{\mathrm{PL}})$} &
\colhead{Index $\Gamma$} &
\colhead{Flux $\log(\Phi_{\mathrm{SE}})$} &
\colhead{$kT$} &
\colhead{Flux $\log(\Phi_{\mathrm{K}\alpha})$}\\
\colhead{} &
\colhead{} &
\colhead{(c s$^{-1}$)} &
\colhead{(erg~cm$^{-2}$~s$^{-1}$)} &
\colhead{} &
\colhead{(erg~cm$^{-2}$~s$^{-1}$)} &
\colhead{(keV)} &
\colhead{(erg~cm$^{-2}$~s$^{-1}$)}
}
\startdata
NICER XTI & 164 & 50.6 & 15.9 & 2.8 & 2.5 & 1.5 & 1.7\\ 
Swift XRT & 156 & 1.1 & 3.0 & 1.1 & 1.3 & 2.2 & 1.8\\ 
\enddata
\tablecomments{Ratios between the standard deviation of best-fit values for each parameter across all epochs ($\sigma_\mathrm{std}$) and the mean parameter error at the 68\% confidence level ($\sigma_\mathrm{mean}$). We consider the spectra considered sensitive to changes in a parameter if $\sigma_\mathrm{std}/\sigma_\mathrm{mean}>2$. All count rates and fluxes cover the 0.3--10 keV band.}
\end{deluxetable*}

\begin{deluxetable*}{cccccc}[t]
\tablewidth{0pt}
\tablecaption{\label{table:errorconstraincomp}Error Constraint Rate}
\tablehead{
\colhead{Instrument} &
\colhead{Power Law} &
\colhead{Power Law} &
\colhead{Soft Excess} &
\colhead{Soft Excess} &
\colhead{Fe K$\alpha$}\\ 
\colhead{} &
\colhead{Flux $\log(\Phi_{\mathrm{PL}})$} &
\colhead{Index $\Gamma$} &
\colhead{Flux $\log(\Phi_{\mathrm{SE}})$} &
\colhead{$kT$} &
\colhead{Flux $\log(\Phi_{\mathrm{K}\alpha})$}\\
\colhead{} &
\colhead{(\%)} &
\colhead{(\%)} &
\colhead{(\%)} &
\colhead{(\%)} &
\colhead{(\%)}
}
\startdata
NICER XTI & 100 & 100 & 100 & 99 & 66 \\ 
Swift XRT &  100 & 91 & 49 & 32 & 12 \\ 
\enddata
\tablecomments{Percentage of epochs for which both the upper and lower $1\sigma$ errors for a given parameter are constrained, e.g. within the allowed limits of Table~\ref{table:modelparams}. Count rates and fluxes cover the 0.3--10 keV band.}
\end{deluxetable*}

\newpage

\section{Optical Cross-Correlation Analysis}
\label{sec:gbandappendix}
\begin{figure*}[t]
    \centering
    \includegraphics[width=0.9\textwidth]{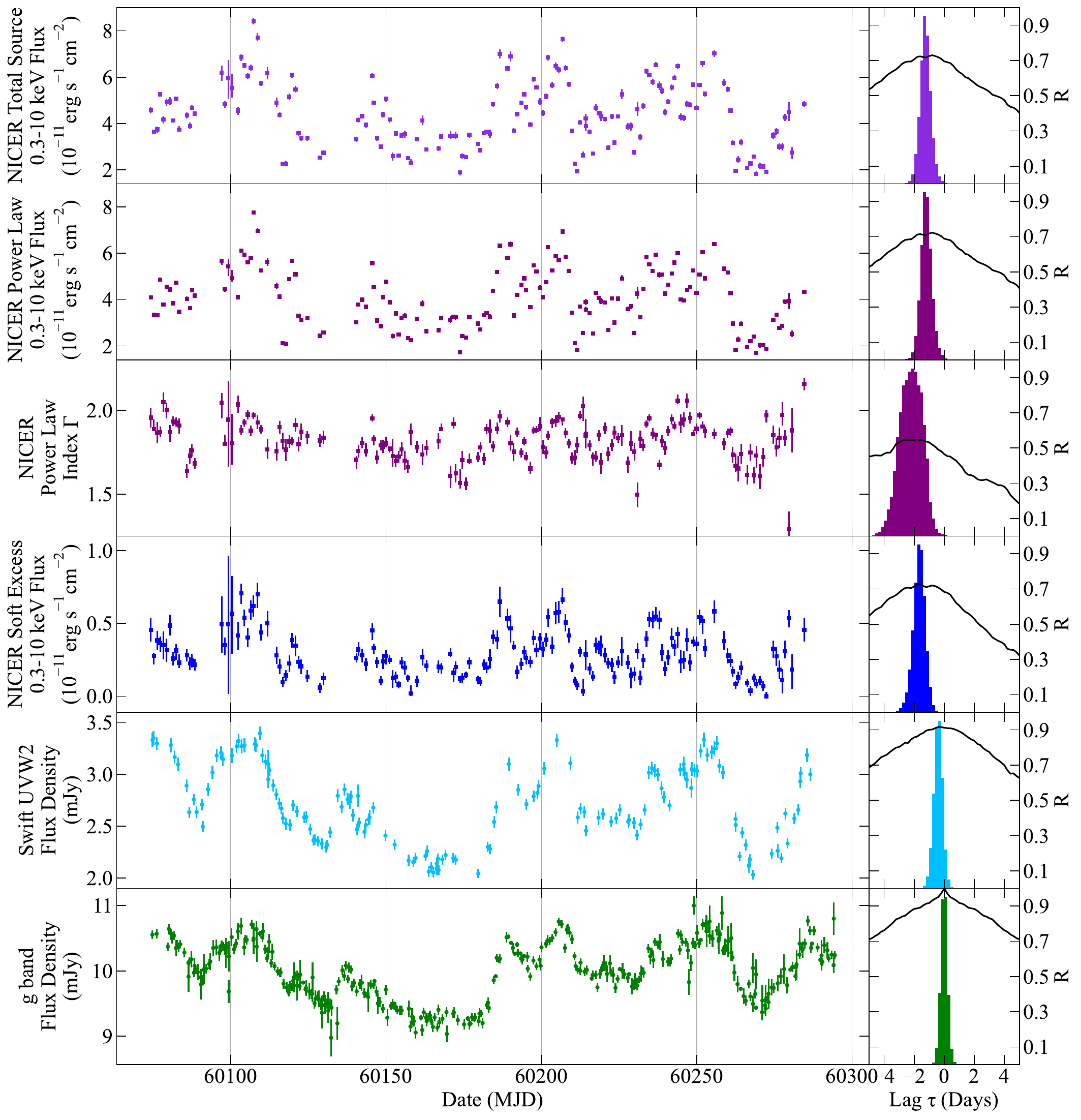}
    \caption{The same as Figure~\ref{fig:nobinCCF}, except for the inclusion of the optical \textit{g}-band light curve (left panel, green squares) which is used as the reference for the ICCF (right panel).}  
\label{fig:gbandccf}
\end{figure*}

Ground-based optical measurements in the \textit{g}-band with LCO and Zowada were taken at a faster cadence than UVW2 monitoring with Swift. This motivates our measurement of the ICCFs of the X-ray and UVW2 light curves relative to the reference \textit{g}-band (Figure~\ref{fig:gbandccf}). The faster observing cadence produces an X-ray/optical lag measurement with a smaller uncertainty ($\pm 0.4$ days, Table~\ref{table:gcentlags}) than the X-ray/UV lag uncertainty ($\pm 0.5$ days, Table~\ref{table:UVcentlags}). However, the X-ray power law/optical correlation ($R_\mathrm{cent}=0.71$) is also lower than the X-ray power law/UV correlation ($R_\mathrm{cent}=0.77$). The \textit{g}-band and UVW2 light curves are highly correlated ($R_\mathrm{cent}=0.92$), with a lag consistent with zero days. The implications of these results are discussed in Sections~\ref{sec:xrayreprodiscuss} and~\ref{sec:geomdiscuss}.

\clearpage
\bibliographystyle{apj}
\bibliography{ref.bib}
\end{document}